\documentclass[12pt]{article}
\usepackage{amssymb,amsmath}
\begin{document}

\begin{titlepage}

                            \begin{center}
                            \vspace*{1cm}
\Large\bf{Groups, non-additive entropy and phase transitions}\\

                            \vspace{2.5cm}

              \normalsize\sf    NIKOS \  \ KALOGEROPOULOS $^\ast$\\

                            \vspace{0.2cm}
                            
 \normalsize\sf Weill Cornell Medical College in Qatar\\
 Education City,  P.O.  Box 24144\\
 Doha, Qatar\\

                            \end{center}

                            \vspace{2.5cm}

                     \centerline{\normalsize\bf Abstract}
                     
                           \vspace{3mm}
                     
\normalsize\rm\setlength{\baselineskip}{18pt} 

\noindent We investigate the possibility of discrete groups  furnishing a kinematic framework
for systems whose thermodynamic behaviour may be given by non-additive entropies. 
Relying on the well-known result of the growth rate of balls of nilpotent groups,  we see  
that maintaining extensivity of the entropy of a nilpotent group requires using a 
non-Boltzmann/Gibbs/Shannon (BGS) entropic form. We use the Tsallis entropy as an 
indicative alternative. Using basic results from hyperbolic and random groups, we investigate 
the genericity and possible range of applicability of the BGS entropy in this 
context.  We propose a sufficient condition for phase transitions, in the context of (multi-) 
parameter families of non-additive entropies.

                             \vfill

\noindent\sf  PACS: \  \  \  \  \  02.20.Bb, \ 02.20.Hj, \ 02.40.Hw, \ 05.70.Fh, \  05.90.+m.  \\
\noindent\sf Keywords:  Tsallis entropy, Non-extensive entropy, Nilpotent groups, Hyperbolic, Phase transitions.   \\
                             
                             \vfill

\noindent\rule{8cm}{0.2mm}\\  
   \noindent $^\ast$ \footnotesize\rm E-mail: \ \  nik2011@qatar-med.cornell.edu\\

\end{titlepage}


                                                                                 \newpage

                                          \normalsize\rm\setlength{\baselineskip}{18pt}

                                                 \centerline{\large\sc 1.   \ Introduction}
                                                 
                                                                                 \vspace{5mm}

In this work we present discrete groups as a kinematical framework for a set of toy-models, 
whose thermodynamic description can  be described by a non-additive entropy.
To be concrete, we use the Havrda-Charvat [1], Dar\'{o}czy [2], Cressie-Read [3], [4], Tsallis [5], [6] (henceforth to be called just ``Tsallis", 
for brevity) entropy as a non-additive generalisation of the Boltzmann/Gibbs/Shannon (BGS) entropy. 
We are partly motivated by the scarcity of analytically tractable examples as toy-models that may be used to explore properties  of the Tsallis 
entropy. One notable example that has been extensively analysed in the context of Tsallis entropy is that of correlated binary systems [7]-[11].         
Aspects of the the analysis of such systems were generalised in [12] where it was argued that use 
of the Tsallis entropy in the statistical description of a system is related to the rate of growth of its configuration/phase volume.  
Such a volume growth function has been  analysed in the context of Riemannian manifolds which form the configuration or phase 
space of many systems of significance in Statistical Physics. Since a dynamical analysis of systems described by flows on Riemannian 
manifolds whose thermodynamic behaviour is described by the  BGS or Tsallis 
entropies is clearly beyond our immediate reach, we have been motivated to find simpler structures where a similar behaviour occurs.
Such a relatively simple, but also quite rich, structure turns out to be that of discrete groups.\\  

Investigations of discrete groups have a long history going back to antiquity. In contemporary Physics, they are used very extensively in a 
wide range of topics from  crystallography [13] to quantum field theory [14]. The more recent theory of discrete groups has been developing \
for almost two centuries, since the time of Galois; we will use its facet initially known  as combinatorial group theory [15], [16]. 
The contributions [17], [18] had a transformative effect on combinatorial group theory, bringing it closer than ever before to geometry, by 
abstracting the large-scale/coarse features of some of these constructions. Hence features 
well-known in Physics from Riemannian geometry can now also be seen in discrete groups following the work of Gromov [17], [18]. 
For simplicity, we will only deal with (discrete) finitely generated and even finitely presented groups in this work. 
We will state some well-known links with the geometry of manifolds, wherever possible, in order to show that many of the underlying 
ideas are actually quite familiar in Physics.\\

The goal of such a ``geometric" group theory [19]  is to explore properties of discrete groups either by examining their action on manifolds 
or metric spaces, or by constructing spaces such as the Cayley graphs, whose geometric features encode algebraic properties of the 
underlying groups [17]-[19]. 
It attempts to infer algebraic properties of the groups by using geometric arguments pertaining to such spaces. 
We  provide the motivation, whenever possible, and present connections of the concepts discussed  
from the  viewpoint of statistical mechanics and classical/quantum field theory. We are rather heuristic than even attempt to be 
rigorous, throughout this work. We  provide a bare minimum of definitions and results of combinatorial/geometric group theory so as to 
allow the reader to follow the line of 
reasoning and its consequences presented here.  For a more extensive background and full justifications / proofs of all statements,
illustrated with numerous examples, we cite a few pertinent references  of  combinatorial/geometric group theory on which we also 
extensive rely.  \\      

                                                                            \vspace{5mm}


  \centerline{\large\sc 2.   \ Discrete groups: pertinent features}
                                                 
                                                                                 \vspace{5mm}

\noindent{\large\bf 2.1} \ \ Groups are, arguably, one of the simplest and most elegant algebraic structures.  A group, is by definition, a set 
\ $\mathcal{G}$ \   endowed with a binary operation  henceforth to be indicated by \ $\cdot$ \ and to be called  ``multiplication" [20]. 
The set \ $\mathcal{G}$ \ should be closed  under multiplication, namely \ $ab\equiv a\cdot b \ \in \mathcal{G}, \  \forall \ a,b \in \mathcal{G}$, 
\ and if \ $c \in \mathcal{G}$ \ then 
\begin{equation}
     a(bc) \ = \ (ab)c     \hspace{20mm}  \mathrm{(Associativity)}
\end{equation}
The structure ($\mathcal{G}, \cdot$) with the binary operation being just associative  is called a semigroup. 
Examples of such a structure are the renormalization ``group" and the heat semigroup, both of considerable importance in Physics. 
If, in addition, there is an element  $1\in\mathcal{G}$ \ such that 
\begin{equation}
    a\cdot 1 \ =  \ 1\cdot a \ = \ a     \hspace{20mm} \mathrm{(Existence \ of  \ unit)}
\end{equation}
then structure  \ ($\mathcal{G}, \cdot$) \ with the binary operation obeying (1), (2) is called a monoid.
Consider a set \ $A =\{ a_i \}_{i\in I}$, \ where \ $I$ \ is an index set, and define the set of finite sequences of elements of \ $A$ \  
denoted by \ $\mathcal{F}(A)$. \ The set \ $A$ \  is called the generating set or the set of generators of \ $\mathcal{F}(A)$. \ 
Elements of \ $\mathcal{F}(A)$ \ are words (formal products) made up of elements  of \ $A$ \ such as 
\begin{equation}
                 w \ = \ a_{i_1}a_{i_2} \ldots a_{i_k}, \hspace{15mm} a_{i_1}, \ldots a_{i_k} \ \in \ A
\end{equation} 
Define as multiplication on \ $\mathcal{F}(A)$ \ the concatenation of words and as the 
identity element of \ $\mathcal{F}(A)$ \ the empty word. Then \ $\mathcal{F}(A)$ \ becomes a monoid and is called the free monoid 
on \ $A$, for obvious reasons. 
If, in addition, for each \ $a\in\mathcal{G}$ \ there is \ $b\in\mathcal{G}$ \ such that 
\begin{equation} 
    ab \ = \ ba \ =  \ 1     \hspace{20mm}  \mathrm{(Inverse \ element)}
\end{equation}
The notation for such \ $b$ \ is \ $a^{-1}$. Then  \ ($\mathcal{G}, \cdot$) \ obeying (1), (2), (4) is called a group. 
If \ $\mathcal{H} \subset \mathcal{G}$ \ is closed under the restriction of the multiplication of \ $\mathcal{G}$, \ then \ $\mathcal{H}$ \ 
is called a subgroup of \ $\mathcal{G}$ \ and it is indicated by \ $\mathcal{H} < \mathcal{G}$.  \  There is also the more restricted 
definition of \ $\mathcal{H}$ \ being a  normal subgroup of \ $\mathcal{G}$: \ this is so, if for all elements \ $h\in\mathcal{H}$ \ 
all its $\mathcal{G}$-conjugates belong to \ $\mathcal{H}$, \ namely \  $ghg^{-1} \in\mathcal{H}, \  \forall \ g\in\mathcal{G}$. \     
We indicate such a  normal subgroup by writing \  $\mathcal{H} \   \lhd  \   \mathcal{G}$. \ So, we see that a subgroup is normal, when it
remains invariant under the action of the whole group by conjugation.
If the set \ $A$ \ is a group, we can extend the free monoid construction of (3) by defining the inverse word \ $w^{-1}$ \ of  \ $w$ \ in (3) by
\begin{equation}
   w^{-1} \ = \ a_{i_k}^{-1} \ldots a_{i_2}^{-1} a_{i_1}^{-1}  
\end{equation}
Consider the  equivalence relation between words \ $w_1, w_2 \in \mathcal{F}(A)$ \ generated by 
\begin{equation} 
    w_11w_2 \ \sim \ w_1w_2
\end{equation}
and
\begin{equation}
 w_1 b_1 b_2 w_2 \ \sim \ w_1 b_3 w_2, \hspace{10mm} \mathrm{if} \ \ \ \ b_3 = b_1 b_2, \ \ \  b_1, b_2, b_3 \in \mathcal{F}(A)
\end{equation} 
Then the set of equivalence classes \ $\mathcal{F}(A) / \sim $ \ is called the free group of \ $A$. \ 
So $\mathcal{F}(A)$ is the set of reduced words in \ $A$.  If \ $A$ \ is not assumed to be a group then \ $\mathcal{F}(A)$ \ is the set of free 
words on \ $A\cup A^{-1}$ \ where \ $A^{-1}$ \ has elements given by (5), after concatenation and reduction (6), (7). 
The cardinality of \ $A$ \ is called the rank of the free group \ $\mathcal{F}(A)$. \ 
Obviously \ $A$ \ is a subgroup of \ $\mathcal{F}(A)$ \ corresponding to single letter words. More generally the length \ $l(w)$ \ of the word 
\ $w\in \mathcal{F}(A)$ \ shown in (3), is the number of elements of \ $A$ \ needed to make up \ $w$. \ If an element of \ $A$ \ appears twice or 
more in the reduced word \ $w$ \ then it contributes two (or more) units to its length.  Obviously \ $l(w)\in\mathbb{N}$. \\ 

The universal property of the free group construction, which makes it particularly useful, is that  any group is the quotient of some free group 
[19], [20]. To construct any other group, one has to start from the free group and provide some additional ``constraints". This set of constraints
is expressed by setting words on \ $\mathcal{F}(A)$ \ equal to identity. Such constraints  are called relations \ $R$. \ 
A group \ $\mathcal{G}$ \ is ``produced" as the  result of identifying the elements of \ $\mathcal{F}(A)$ \ that belong to the normal 
closure of \ $R$. \  This group specification by its 
generators \ $A$ \ and its relations \ $R$, \ is called a presentation of \ $\mathcal{G}$ \ and is indicated by \ $\langle A|R \rangle $. \ 
If \ $\mathcal{G}$ \ admits a presentation for which the cardinality of \ $A$ \  is finite then the group is called finitely generated. 
If the same applies to both \ $A$ \ and \ $R$ \ then the group \ $\mathcal{G}$ \ is called finitely presented. It should be noticed that 
the presentation of a a group is not unique. It should also be noticed that, somewhat unexpectedly for such  ``innocent" looking 
definitions, it turns out that answering whether a given presentation determines the trivial group or not, 
is an algorithmically undecidable proposition [17], [19].\\

The specification of discrete groups via presentations may appear familiar in Physics. Indeed consider the most familiar 
way that a Classical or Quantum theory is constructed. Usually one starts with a set of fields which are sufficiently smooth (or integrable)
maps  from space(-time) to some space. Such a target space encodes the ``internal"/isospin  degrees of freedom and their symmetries. 
Then one   expresses the dynamics of the model, frequently in the Hamiltonian approach and following Dirac, through some 
``constraints" [21]: the equations associated with the time-evolution of the system provide the dynamical equations for the fields of interest 
and the spatial equations provide the constraints of the system, in this particular approach.  \\

From the viewpoint of large-scale properties of groups, picking a group presentation may be seen as a very rough analogue,  
to choosing a particular parametrization for a physical model. The large-scale features of the group should not depend on the presentation as 
much as the physical results that the model predicts cannot depend on the parametrization (``gauge") used to reach them. \\  

For completeness, we this section finish by mentioning that if for all \ $a,c \in \mathcal{G}$
\begin{equation} 
   ac \ = \ ca    \hspace{20mm} \mathrm{(Commutativity)}
\end{equation}   
then the group is called commutative  or Abelian. See [20], [19] for more details on these matters.\\


\noindent{\large\bf 2.2} \ \ One can always wonder, and even object, about the use of groups in our context. 
From a physical viewpoint, groups have been used  to quantify the concept 
of symmetry. This is clearly seen in various physics sub-disciplines. In Particle Physics, in particular, symmetry expressed through groups,
has been elevated to be the guiding principle (the ``gauge principle") behind the construction of models of fundamental interactions.
Similar considerations apply to General Relativity and to the conjectural approaches to quantum gravity such as geometrodynamics, loop 
gravity, causal sets etc, as well as to the unification of interactions attempted by such approaches as string/brane/M-theories, supergravity etc. 
In Statistical Mechanics proper, the use of groups occupies somewhat less of a central role, but it still remains an important concept 
in the theory of phase transitions (for instance), not to mention its use in simplifying expressions when performing explicit calculations.\\ 
 
For our purposes, groups are useful because they are simple enough to allow inference of non-trivial results  and because they 
also involve a rudimentary concept of dynamics. Indeed, someone can interpret the multiplication in a group as expressing the
composition properties of the system it describes, which is quite important in determining its dynamics. Consider, by contrast, still for our 
purposes only, binary systems [7]-[11]: although they may be, arguably, the simplest systems in discussing the meaning of  
``independence" and the the applicability of the  BGS or the Tsallis entropies, they have ``too little" intrinsic structure to allow even 
rudimentary dynamical considerations to be developed: such dynamics has to be additionally 
imposed as correlations, see for instance [7]-[11], often having desirable singularity and/or asymptotic behaviours. 
Groups have more structure than binary systems, thus 
allowing aspects of some dynamical features to be incorporated via their multiplication, as will be seen in the sequel. 
Even though will be discussing discrete groups only,  some of the quoted results can be carried over to ``continuous" groups.\\

Another reason for using groups is the following: even though it is not obvious that most configuration of phase spaces of the systems of 
interest have an underlying group theoretical structure, they may be either reduced to such structures or may be effectively approximated by 
such structures in limiting behaviours  (such as the thermodynamic limit) of interest. 
An exemplary approach of such a reduction is the topological reduction of 3-manifolds after a prime and a Jaco-Shalen/Johansson (JSJ) 
decomposition into the eight possible geometries first classified by Thurston [22]. Each of these geometries 
has as cover one of the following eight groups or their homogeneous spaces: \ $\mathbb{E}^3$ \ (Euclidean), \ $\mathbb{H}^3$ \ (hyperbolic),
\ $\mathbb{S}^3$ \ (spherical), \ $\mathbb{S}^2 \times \mathbb{R}$, \ $\mathbb{H}^2 \times \mathbb{R}$, \ $\widetilde{SL_2(\mathbb{R})}$ \ 
(universal cover of $SL_2(\mathbb{R} )$), \ Nilpotent and Solvable. In addition, is true that if a 3-manlfold admits a geometric structure, then 
such a geometric structure is unique. Such a classification is obviously of great interest to General Relativity, especially if one discusses 
globally hyperbolic (in the Lorentzian sense) space-times where the initial data are specified on a space-like hyper-surface, where global 
hyperbolicity is needed so  one can develop a consistent Hamiltonian formalism [23]. Something similar can be stated about classical and 
quantum field theories with scalars, in the perturbative regime at least [24]: their vacua can be expressed as homogeneous spaces of the 
symmetry group in the global 
symmetry case, or as homogeneous spaces of a local section of the gauge group in the case of ``local" symmetries [25]. The groups that we 
are referring to in this paragraph are either Lie groups or more general topological groups. So they may appear to be very far removed from 
the discrete groups that this work is all about. However such distinction may become less relevant if one is willing to see such spaces 
coarsely-grained, something which is metrically implemented by requiring invariance under quasi-isometries of the structures of physical 
interest. Besides, discrete groups can appear in this context in case of discretizations  of the underlying groups useful for regularization 
purposes or in numerical simulations. \\ 

It is unlikely that statements such as the geometric classification of 3-manifolds can possibly  be made for the high dimensional manifolds,
reflecting the very high number of degrees of freedom, that serve as configuration or phase spaces of the systems of interest in Statistical 
Mechanics. The question of the existence of such structures may even lead to undecidable propositions. However, for the purposes of 
Statistical Mechanics, it is sufficient if some key features of these underlying configuration/phase spaces can have a group theoretical 
description. These key features should be the ones determining the thermodynamic behaviour of the underlying system.
A motivational analogy can be borrowed from the chaotic hypothesis  [26], [27]    according to which a system having chaotic motions 
does it in a maximal way so that it can be approximated as a transitive hyperbolic (in the dynamical sense) system. Here the underlying 
system does not have to be transitive hyperbolic but to be approximated by one for all intends and purposes of interest. By the same token,
a physical system described the BGS, the Tsallis or any other entropic form does not have to have a discrete  underlying group theoretical 
structure, as long as it can be sufficiently well approximated by one, in the thermodynamic limit. Configuration spaces of lattice models, 
such as the Ising, the Potts, the spherical etc   possess such group theoretical structures. Similar to the chaotic hypothesis, one encounters 
such an approximation in the general theory of dynamical systems under the label of ``shadowing" [28]. \\  
 
   
\noindent{\large\bf 2.3} \ \ As mentioned above, a group \ $\mathcal{G}$ \ is finitely generated if there is a finite subset 
\ $\Gamma \subset \mathcal{G}$ \ such that for any \ $g\in\mathcal{G}$ \ there are \ $\gamma_1, \ldots, \gamma_n \in \Gamma$ \ 
such that \ $g = \gamma_1\cdots \gamma_n$. \ In other words,    
such  \ $\mathcal{G}$ \ is the quotient of a free group of finite rank. It is immediate to see that finitely generated groups are countable. 
Somewhat conversely, it may not be so easy to decide on whether a countable group is finitely generated or not. 
Unexpectedly,  geometry can help in this respect, by seeing such  group as a metric space thorough its Cayley graph [17], [18]. 
This will be discussed in next paragraphs. Notice also that every countable group can be embedded in a group with two generators. 
Examples of  finitely generated groups are \ $\mathbb{Z}^n$ \ or the general linear group \  $GL(n,\mathbb{Z})$ \ for \ $n\in\mathbb{N}$. \  
By contrast, an example of a non-finitely generated group is the multiplicative subgroup of the rationals \ $\mathbb{Q}_\ast \subset \mathbb{Q}$ \ 
or the special linear group \ $SL(n,\mathbb{K})$ \ over a field \ $\mathbb{K}$ \ of zero characteristic.\\

Suppose that \ $A$ \ is a finite symmetric set and \ $\mathcal{F}(A)$ \ is the free group of \ $A$. \ A finitely generated 
group \ $\mathcal{G}$ \ can be seen as a quotient \ $\pi: \mathcal{F}(A) \rightarrow \mathcal{G}$. \ Then any element \ $g$ \ of \ $\mathcal{G}$ \ 
can be written as 
\begin{equation}
 g = \pi (\gamma_1 \cdots \gamma_n), \hspace{5mm} \gamma_i \in A, \hspace{5mm}  i=1\ldots, n 
\end{equation}
We usually identify, abusively, \ $A$ \ and its image in \ $\mathcal{G}$. \  
The word length \ $l_A(g)$ \ of such a \ $g\in\mathcal{G}$ \ is the smallest \ $n\in\mathbb{N}$ \ for which such a word (9) exists. 
The word metric \ $d_A (g, h), \   g,h \in\mathcal{G}$ \ is defined as 
\begin{equation}    
      d_A(g,h) \ = \ l_A (h^{-1}g)
\end{equation} 
Then \ $\mathcal{G}$ \ endowed with the word metric \ $d_A$ \ becomes a metric space. Obviously \ $\mathcal{G}$ \ acts upon itself 
from the left by isometries. To visualise this construction one can use a regular graph, called the Cayley graph 
\ $\Gamma_A (\mathcal{G})$ \ of \ $\mathcal{G}$ \ 
with respect to the generating set \ $A$. \ This is an un-oriented graph \ $\Gamma_A (\mathcal{G})$ \ which is defined as having vertices the 
elements of  \ $\mathcal{G}$. \ Two vertices \ $g, h$ \  of \ $\Gamma_A (\mathcal{G})$ \ are connected by an edge if their word distance is \ 
$d_A(g,h) = 1$. \  Then clearly \ $\Gamma_A (\mathcal{G})$ \ is a path-connected space with the embedding \ $\mathcal{G} \hookrightarrow 
\Gamma_A (\mathcal{G})$ \ being an isometry.  The Cayley graph is a homogeneous space: indeed any vertex of the graph can be used 
as the identity element of the group. For finitely generated groups, their Cayley graph is a proper metric space, namely the closed balls are 
compact [19].\\

 The Cayley graph \ $\Gamma_A (\mathcal{G})$ \ strongly depends on the choice of the generating set $A$
and, as such, it may not appear to be a good space to use to explore intrinsic (algebraic) properties of \ $\mathcal{G}$. \ Changing 
the generating set of a group, may give rise to even non-homeomorphic Cayley graphs [19]. Things can get even 
more complicated by that fact that different groups can have isomorphic Cayley graphs. However the coarse features of Cayley graphs of 
groups, encoded in their quasi-isometry classes depend only on \ $\mathcal{G}$, \  and are independent of \ $A$ \ as will be seen in the sequel.
Some examples of Cayley graphs are: for the free group of \ $m$ \ generators, its Cayley graph is an infinite tree of valence \ $2m$. \ 
The Cayley graph of the cyclic group of order $n$, \ $\mathbb{Z}_n$ \ is an $n$-(poly)gon. The Cayley graph of the group \
$\mathbb{Z}^2 = \mathbb{Z} \times \mathbb{Z}$ \ is a 
regular lattice whose vertices have integer coordinates on \ $\mathbb{R}^2$ \ and its edges connect vertices with one unit of difference in their 
coordinates. This Cayley graph looks like a grid of horizontal and vertical lines intersecting at points having both points with integral 
coordinates [17], [19].\\  
    

\noindent{\large\bf 2.4} \ \ Since Cayley graphs endowed with the word metric (10) are metric spaces, it may be worth examining when 
they are ``equivalent".
Probably the simplest metric equivalence relation between metric spaces is that of isometry. 
Consider two metric spaces  ($X_1, d_1$)  and ($X_2, d_2$) to be two metric spaces with corresponding distance functions
 \ $d_1, \ d_2$. \  A map \ $f: X_1 \rightarrow X_2$ \ is an isometry if it is surjective 
\begin{equation}
      f(X_1) \ =  \ X_2
\end{equation}
and distance-preserving 
\begin{equation}
     d_2 (f(x), f(y)) \ = \ d_1 (x,y), \hspace{10mm}  \forall \  x,y \in X_1 
\end{equation} 
This definition is probably the most used one in geometry, but has its inevitable limitations. One of them is that it is too restrictive when one is 
interested only in ``large-scale" or ``asymptotic" metric properties. The latter are very useful in Statistical Physics as they may encode aspects 
of the system that still hold in the thermodynamic limit. Looking at a metric space from such a ``coarse" viewpoint, inevitably eliminates its 
small-scale features. On the other hand it allows spaces that are very different, in the small, but mostly the same asymptotically, to be treated 
in a unified manner. One such space can be discrete, for instance, and the other can be continuous. 
For such a comparison, the concept of quasi-isometry is widely used. 
Consider two metric spaces ($X_1, d_1$), ($X_2, d_2$) as above. Let \ $L\geq 1, \ C>0$ \ be two constants. A map 
\ $\varphi : X_1 \rightarrow X_2$ \ is an \ ($L, C$) \  quasi-isometry if it is almost surjective, namely if there is a constant \ $A\geq 0$ \ such that 
\begin{equation}    
        d_2 (\varphi (X_1), X_2) < A
\end{equation}
and if 
\begin{equation}
     \frac{1}{L} \ d_1 (x,y) - C \ \leq \ d_2 (\varphi(x), \varphi(y)) \ \leq \ L \ d_1(x,y) + C  
\end{equation}
There is no requirement for \ $\varphi$ \ to be continuous. In (14), the role of \ $C$ \ is to disregard the small-scale metric structures,
and \ $L$ \ provides  uniform bounds for the amount of distortion that \ $\varphi$ \ can induce. It is obvious, for instance, that any 
bounded space is quasi-isometric to a point. Similarly, a finite group is quasi-isometric to the group with one element. 
This can be re-expressed in terms of Cayley graphs as stating that \ $\Gamma_A (\mathcal{G})$ \ is quasi-isometric to a point 
if and only if \  $\mathcal{G}$ \ is finite.\\ 

It is a non-trivial fact that for a finitely generated group the word metric is unique up to quasi-isometry [17], [19].  
This is a particularly nice property of finitely generated groups as it does not depend on the generating set \ $A$ \ which is usually 
arbitrarily chosen. 
Consider a compact Riemannian manifold \ $\mathcal{M}$, \ such as the configuration space of a bounded system. Then its fundamental 
group \ $\pi_1(\mathcal{M})$ \  is quasi-isometric to \ $\mathcal{M}$ [29], [30]. \ This, in effect, means that when the small-scale details of 
a manifold are ignored and distances can only be distorted by at most a constant, then the manifold is metrically indistinguishable from its 
fundamental group, which is discrete. It provides a substantial connection between properties of discrete groups and  properties of  
metric spaces on which such groups act. There are occasions in which quasi-isometries essentially reduce to isometries. 
This happens when there is considerable  rigidity in the metric structure. A prominent example is that of  irreducible Riemannian symmetric 
spaces of non-compact type [31], 
whose rigidity have made them a very fertile testing ground and sources of examples for numerous ideas.\\

Let \ $\mathcal{H} < \mathcal{G}$. \ Then the index of \ $\mathcal{H}$ \ in \ $\mathcal{G}$ \ is an integer expressing the ``relative size" of 
\ $\mathcal{H}$ \ and \ $\mathcal{G}$. \ More concretely, it counts how many times a copy of \ $\mathcal{H}$ \ fits inside \ $\mathcal{G}$. \ 
Formally, the index is defined to be the number of cosets of \ $\mathcal{H}$ \ in \ $\mathcal{G}$. \ If, for instance, both \ $\mathcal{H}$ \  and 
\ $\mathcal{G}$ \ are finite then their index is 
\begin{equation}
              |\mathcal{G} : \mathcal{H} |  \ =  \   \frac{\mathrm{card} \ \mathcal{G}}{\mathrm{card} \ \mathcal{H}}                 \nonumber
\end{equation}
where \ $\mathrm{card}$ \ stands for ``cardinality". This, intuitively clear, theorem is due to Lagrange [20].
 A group has property ``virtually" P, if it has  subgroup of finite index having the same property. A finite group is virtually 
trivial, for instance. It is probably intuitively straightforward to see that any group quasi-isometric 
to \ $\mathbb{R}^n$ \ is virtually \ $\mathbb{Z}^n$, for instance. Something similar can be stated about metric trees: any group 
quasi-isometric to a tree is virtually free.\\ 

It is straightforward to see that for the Euclidean 
spaces \ $\mathbb{R}^m$ \ is quasi-isometric to \ $\mathbb{R}^n$ \ if and only if \ $m=n$. \ The same is true about hyperbolic spaces: \
the hyperbolic space $\mathbb{H}^m$ \ is quasi-isometric to \ $\mathbb{H}^n$ \ if and only if \  $m=n$. \ Moreover, the Euclidean spaces   
\ $\mathbb{R}^n$ \ and the hyperbolic spaces \ $\mathbf{H}^n$ \ are not quasi-isometric to each other for any \ $n, m \geq 2$. \ 
This should be obvious as Euclidean distances are expanding exponentially, i.e. unboundedly thus violating (14),  in hyperbolic spaces. 
In addition it may be worth observing that any two word metrics on a given group are quasi-isometric (actually, they are ``bi-Lipschitz 
equivalent", which is (14) with $C=0$) to each other. In terms of groups, if $\mathcal{F}_m$ is a free group on $m$ generators, we observe that 
all such \ $\mathcal{F}_m, \ m\in\mathbb{N}$ \ are quasi-isometric to each other. By stark contrast, this is not true for groups of polynomial growth 
(for the definition see the next section).  As an example, \ $\mathbb{Z}^m$ \ and \ $\mathbb{Z}^n$ \ are not quasi-isometric to each other 
for \ $m\neq n$. \\


\noindent{\large\bf 2.5} \ \ Consider a discrete group \  $\mathcal{G} $ \ with a symmetric generating set \ $A$ \ equipped with the word metric 
\ $d_A$ \ (10). For \ $n\in\mathbb{N}$, \ consider the closed ball  
\begin{equation}
   \bar{B}_g (n) \ = \ \{ h\in \mathcal{G} :  \  \  d_A (g,h) \leq n \}
\end{equation}
Let \ $\beta (n)$ \ indicate the cardinality of \ $\bar{B}_g (n)$ \ which is clearly independent of \ $g\in\mathcal{G}$ \ due to that the left action of 
\ $\mathcal{G}$  \ upon itself is an isometry, as also noticed above. As a result, instead of \ $\beta_g(n)$ \ one can write just \ $\beta (n)$. \ 
Then \ $\mathcal{G}$ \ 
has polynomial growth is there are constants \ $c_1>0, \ c_2, \ d>0$ \ such that 
\begin{equation}
     c_1n^d   \leq \ \beta (n) \ \leq \ c_2 n^d,  \ \ \ \ \forall \  n \in  \mathbb{N}
\end{equation} 
A group has exponential growth if for some \ $\alpha > 1$ 
\begin{equation}
       \beta (n) \ \geq \   \alpha^n
\end{equation}
and has sub-exponential growth otherwise. If a group has a growth which is sub-exponential but non-polynomial then it is called 
a group of intermediate growth. It is straightforward to see that for a finitely generated group of rank \ $k$ \ the growth function is maximal
for the free group on $k$ generators which has exponential growth 
\begin{equation}
   \beta (n) \  =  \ 2k (2k-1)^{n-1}
\end{equation}
More generally, if a group has a free subgroup of even two generators, then it has polynomial growth. 
By contrast, finite groups and free Abelian groups have polynomial growth. 
In particular, the Abelian group \ $\mathbb{Z}^k$ \ has polynomial growth of degree $k$ \ [17]-[19].   \\

As can be readily seen by (15),  the growth function of a group is the analogue of the volume growth of a  
Riemannian manifold. Given some of the above, not too surprisingly, there are strong relations between the volume growth of a 
Riemannian manifold, with appropriate curvature bounds, and  the growth of its fundamental group: in particular, the growth type of the 
fundamental group of a Riemannian manifold endowed with the word metric is equal to the growth rate of volumes in its universal cover [17]. 
It is non-trivial fact that the growth of a finite generated group is invariant under quasi-isometries of the corresponding Cayley graphs [17], [19]. \\

  
\noindent{\large\bf 2.6} \ \ Let \ $\mathcal{G}$ \ be a group and \ $\mathcal{H} < \mathcal{G}$. \ Then the commutator 
\ $[\mathcal{H}, \mathcal{G}]$ \ of \ $\mathcal{G}$ \ and  \ $\mathcal{H}$ \ is defined as the group generated by  
\begin{equation}
    [\mathcal{H}, \mathcal{G}] \ = \  \langle \ h^{-1}g^{-1}hg, \ \ \    g\in\mathcal{G}, \ h \in\mathcal{H} \ \rangle  
\end{equation}
Consider the sequence of successive commutators
\begin{equation} 
   \mathcal{G} = \mathcal{G}_0, \ \ \ \  \mathcal{G}_1 = [\mathcal{G}, \mathcal{G}], \ \ \ \  \mathcal{G}_2 = [\mathcal{G}_1, \mathcal{G}], \ \ \ \  
             \mathcal{G}_3 = [\mathcal{G}_2, \mathcal{G}], \ \ \ \ \ldots
\end{equation}
Observing that \ $\mathcal{G}_{j+1} < \mathcal{G}_{j}, \ \ j = 0,1,2, \ldots$ \ the sequence of commutators can be organized in what is 
called the lower central series   
\begin{equation}
     \mathcal{G}_0 \   >  \  \mathcal{G}_1 \   >  \  \mathcal{G}_2  \  >  \  \mathcal{G}_3 \  > \  \ldots
\end{equation}
 If (20) terminates after a finite number of steps then \ $\mathcal{G}$ \ is called nilpotent. Obviously,  Abelian groups is nilpotent.
 A nilpotent, non-Abelian group of great significance in Physics is the Heisenberg group.  
The significance of nilpotent groups is that  they provide is a generalisation of  Abelian groups but also obey a similar to the Abelian
finiteness condition given by (21). As metric spaces, nilpotent groups are equivalent to almost flat manifolds, namely to Riemannian manifolds   
with almost vanishing sectional curvature [32], [33].\\
 
                                                                             \vspace{5mm}


\centerline{\large\sc 3. \ Tsallis entropy, volume growth and nilpotent groups}

                                                                         \vspace{5mm}

\noindent{\large\bf 3.1} \ \ The Havrda-Charv\'{a}t [1], Dar\'{o}czy [2], Cressie-Read [3], [4], Tsallis [5], [6] 
entropy is a one-parameter family of functionals that has attracted attention 
during the last twenty five years, and has become the basis of a generalised thermodynamics. 
For a system, with a discrete set of outcome probabilities \ $\{ p_i \}, \ i\in I$ \ labelled by the index set \ $I$ \ the 
Tsallis entropy is   
\begin{equation}
     S_q [\{ p_i \} ] \ = \ k_B \ \frac{1}{q-1} \left(1 - \sum_{i\in I} p_i^q \right)
\end{equation}
where we will set from now on the Boltzmann constant \ $k_B=1$, \ for simplicity, and where \ $q\in\mathbb{R}$ \ is called entropic or 
 non-extensive parameter. The Tsallis entropy reduces to the BGS entropy, namely 
\begin{equation}
      \lim_{q\rightarrow 1} S_q \ = \ S_{BGS}
\end{equation}
as can be readily seen.  
Consider two systems \ ${\sf A, B}$ \ with corresponding probabilities of  occurrence \ $p_{\sf A}, \ p_{\sf B}$. \ Then ${\sf A, B}$ \  
are independent, if the probability of the combined system ${\sf A \cup B}$ obeys \ $p_{\sf A\cup B} \ = \ p_{\sf A} p_{\sf B}$. \ For two such 
independent systems the Tsallis entropy is non-additive, since 
\begin{equation}   
      S_q [\{ p_{\sf A \cup B} \} ] \ = \ S_q [\{ p_{\sf A} \} ] + S_q [\{ p_{\sf B} \} ] + \ (1-q) \  S_q [\{ p_{\sf A} \} ] \  S_q [\{ p_{\sf B} \} ]  \nonumber
\end{equation}

One of the important and still unsolved problems about the Tsallis entropy pertains to the scope of its applicability [6], [34]. 
In particular, what are the common features of the systems whose thermodynamic behaviour is  described by the Tsallis entropy? 
Conjectures and some, largely numerical, evidence certainly abound: systems having long-range interactions, with memory, 
having long-range correlations, described by probability distributions with ``fat tails" etc [6]. To test some these hypotheses, 
toy models have been employed. A notable example is that of a binary system with correlations having specific properties  [7]-[11]. 
This example suggested that systems described by the Tsallis entropy may have phase space volume that does not increase 
exponentially, but rather in a power-law fashion, as a function of the degrees of freedom of the system [7]. 
This set of ideas was further developed in a more general setting in [12] which determined, on quite general
grounds,  that generalized entropies are applicable for systems whose phase space volume increases in a power-law manner 
as a function of the system size, for instance. This  work's [12] paradigmatic examples were binary systems and their $m$-state 
generalizations. Since there are very few analytically tractable examples on which the results of [12] may be explicitly verified, 
we suggest that it may be worth looking into discrete groups  for  constructing toy models of possible interest.\\

The use of discrete groups has its inevitable limitations.  Assume, for instance, that a dynamical system has as  configuration or 
phase space a Riemannian manifold whose effective metric gives rise to a super-exponential deviation of its nearby geodesics. 
Then its greatest Lyapunov exponent would be 
infinite [28]. Such systems however do not seem to be of particular interest for non-additive entropy, as all indications point toward their 
geodesic flows having  strong mixing properties [28]. Such systems are very well described the the BGS entropy. 
The configuraion/phase space behaviour of such systems cannot be described in terms of a group theoretical model 
as the groups with the largest  growth function are the free ones, and their volume growth rate  is only exponential [17]-[19] 
as can be seen in (18).\\

It is the exact opposite side, so to speak, that seems to be of the most interest for non-additive entropies: the case in which the system's
phase space volume increases sub-exponentially [12]. Then groups may be of interest for constructing toy models in such cases. 
An important example of groups with polynomial growth rate are nilpotent groups, such as the Heisenberg group which was mentioned 
above. It turns out that their volume growth function \ $\beta_n$ \ is  polynomial. Such nilpotent groups are  not just important examples,
but it was proved that  these are the only possible cases of groups having polynomial growth functions as will be discussed in 
the next sections.\\


\noindent{\large\bf 3.2} \ \ This fundamental result was obtained over a course of several years. Initially,  Dixmier [35] proved that for a nilpotent 
connected Lie group \ $\mathcal{G}$ \ there exists an integer \ $d(\mathcal{G})$ \ such that given a Haar measure \ $\mu$ \ on \ $\mathcal{G}$ \  
and a compact subset \ $U$ \ of \ $\mathcal{G}$ \ one has \ $\mu(U^n)$ \ is bounded by \ $n^d$ \ as \ $n\rightarrow\infty$. \   
Wolf [36] and Milmor [37] independently proved 
that if \ $\mathcal{G}$ \  is a solvable finitely generated group and has a polynomial growth then it is virtually nilpotent. 
Subsequently Tits [38] proved the ``Tits alternative":  if \ $\mathcal{G}$ \    
is finitely generated subgroup of a linear (Lie) group with finitely many components, then it either contains a free subgroup, in which 
case it has exponential growth, or it is virtually solvable. In the latter case, if one adds the requirement of polynomial growth, the group is 
virtually nilpotent. Guivarc'h [39], [40]  and Bass [41] calculated the degree of growth \ $d$ \ of nilpotent groups in (16), which coincides with their 
homogeneous dimension, and turned out to be 
\begin{equation}
          d \ = \ \sum_{j=0}^{N-1} \ (j+1) \ \mathrm{rank}_\mathbb{Q} \mathbb{G}_j / \mathbb{G}_{j+1}        
\end{equation}
where the notation is that of (19)-(21), \ $N$ \  is the index of the final, hence trivial, commutator in (20) for \ $\mathcal{G}$ \ nilpotent. 
This set of ideas was completed with the proof of Gromov [42] about the converse:  if \ $\mathcal{G}$ \ is finitely generated and has 
polynomial growth then it is virtually nilpotent. More recently, proofs of Gromov' theorem using approaches with an analytical flavor  
have also recently appeared in  [43], [44]. \\

For \ $\mathcal{G}$ \ finitely generated virtually nilpotent, the bound (16) can be strengthened giving rise to the  equality [45]
\begin{equation}
    c \ = \ \lim_{n\rightarrow\infty} \frac{\beta(n)}{n^d}
\end{equation}
There are also some more recent results about the rate of convergence of  such a group to its asymptotic cone [17], [18], such as those of 
[46], [47]. Due to the above volume growth theorem, nilpotent groups may provide an alternative kinematic framework for explicit 
examples extending the validity of some of the results of [7]-[11], [12] in a more general setting. \\


\noindent{\large\bf 3.3} \ \ At this point, it may be worth mentioning the discrete group-theoretical analogue 
of manifolds of negative sectional curvature whose geodesic flow provides some of the best understood cases of chaotic systems [28]. 
Such manifolds are configuration/phase spaces of systems whose thermodynamic properties  are very 
well-described by the BGS entropy. Actually, the following constructions 
are largely motivated by properties of the fundamental group of compact manifolds of negative sectional curvature. 
The pertinent concept is that of hyperbolic groups [17], [18] and their generalisations [18], [48]-[50]. 
Consider \ $\mathcal{G}$ \ to be a finitely  generated group with generating set \ $A$ \ endowed with the word metric (10). 
Define the symmetric bilinear form (Gromov product)
\begin{equation}   
   (g, h) \ = \ \frac{1}{2} \left(l_A(g) + l_A(h) - l_A(g^{-1}h)\right)
\end{equation}
for all \ $g, h \ \in\mathcal{G}$. \ The Gromov product quantifies the idea of the triangle side deficit, or how far is the triangle 
inequality from becoming an equality. In other words, how far is a triangle in a metric space from being a tripod.
Then \ $\mathcal{G}$ \ is a hyperbolic group if there is a constant \ $\varepsilon>0$ \ such that every triple of  
\ $g,h,u\ \in\mathbb{G}$ \ satisfy the inequality
\begin{equation}
    (g, h) \ \geq \ \min \{ (g, u), (h, u) \} - \varepsilon 
\end{equation} 
It should be noted that \ $\varepsilon$ \ is independent of the generating set \ $A$. \  This becomes clearer when re-formulated 
for the more specific case of geodesic metric spaces: a geodesic metric space is $\delta$-hyperbolic if for any geodesic triangle in it, 
each side is contained in a $\delta$-neighborhood of the union of the other two sides. 
This definition expresses the fact that the Cayley graph \ $\Gamma_A(\mathcal{G})$ \ is a hyperbolic metric space, 
namely all its triangles are ``thinner" when compared to triangles in Euclidean space with sides of equal length one-to-one. 
Hyperbolic groups are therefore the group-theoretical analogues of Riemannian manifolds of negative sectional curvature [17].\\ 

Clearly, a finite group is hyperbolic. The same applies to a free group: its Cayley graph is a tree, which  is $0$-hyperbolic.
From a more general, metric viewpoint, all hyperbolic groups look from a large-scale viewpoint as
trees. This can be made precise by using the asymptotic cone of a metric space [18]. indeed, one can turn the arguments 
around and define a finitely generated group to be hyperbolic if its asymptotic cone is a topological tree.   
Some more, equivalent in some settings,  definitions of hyperbolic groups exist [17], [18], 
but the above are sufficient for our purposes. \\


\noindent{\large\bf 3.4} \ \  One very useful  quantity in the geometry of manifolds of non-positive sectional curvature, and in dynamical systems 
in general, is the concept of the volume entropy [28]. Consider a manifold \ $M$ \ and its universal cover 
\ $\widetilde{M}$ \ and let \ $B_x (r)$ \ indicate the open ball of radius \ $r$ \ at \ $x \in \widetilde{M}$. \ 
Then the volume entropy of \ $M$ \ is defined as
\begin{equation}     
    \mathcal{V} (M) \ = \ \lim_{r\rightarrow\infty} \frac{\log vol \ B_x(r)}{r}
\end{equation}
This name is obviously borrowed from the logarithmic character of \ $S_{BGS}$. \ It is straightforward to show that the value of \ 
$\mathcal{V}(M)$ \   is independent of the base point \ $x\in\widetilde{M}$. \  As a example, consider \ $M$ \  to be $n$-dimensional of 
negative constant sectional curvature \ $-k^2$. \ Then 
\begin{equation}
\mathcal{V}(M) \ = \ (n-1) k 
\end{equation}    
According to the above \ $\mathcal{V}(M) >0$ \ if and only if \ $\pi_1(M)$ \ has exponential growth, with respect to any set of generators. 
The volume growth is used as a very rough measure of complexity 
of the geodesic flow on \ $M$. \ In general, the topological entropy of such a flow has a lower bound which is provided by the volume entropy.
When a manifold has non-positive curvature or, more generally, when \ $\pi_1(M)$ \ has exponential 
growth the topological entropy of the geodesic flow and the volume entropy are equal [51].  
Moreover, there is a remarkable rigidity associated with the volume entropy: Consider two compact $n$-manifolds, with \ $n>2$, \  
of negative sectional curvature, one of which is a locally symmetric space. If these manifolds are homotopically equivalent, 
have equal volumes and equal volume entropies, then they are isometric [52].\\

A question of possible interest is to see if there is a corresponding volume-like entropy that is inspired from the Tsallis functional.  From the 
above considerations, it appears that nilpotent groups and almost flat manifolds may provide the general framework. The 
obvious impediment in this endeavor, is that the geodesic flow on almost flat manifolds is far less understood than that on manifolds 
of negative sectional curvature. In the latter case the dynamics is described by a decomposition of the tangent bundle into stable and 
unstable parts both of which are preserved by the geodesic flow [28]. It is not clear even how to formulate, let alone establish, 
such a decomposition for the almost flat manifold case, or if it exists at all. 
At least one can state with some reasonable degree of certainty that the geodesic flow of almost flat manifolds or of manifolds 
with almost negative Ricci curvature, appears to be  a natural geometric ground for searching for the dynamical origins of the Tsallis entropy. \\

                                                                        \vspace{5mm}


\centerline{\large\sc 4. \ A conjecture for phase transitions}

                                                                         \vspace{5mm}

\noindent{\large\bf 4.1} \ \ Phase transitions [53], [54]  are some of the most intriguing phenomena in Physics. Given their theoretical 
importance and physical ubiquity and despite considerable progress,  there are several aspects of them still not particularly well-understood. 
One of the fundamental questions is how can someone predict, starting from first principles (given the Hamiltonian of a system, for instance) 
on whether it will undergo phase transitions, and if so, then what are the features of such a transition. Phenomenological or mean-field  
approaches such as the Landau-Ginzburg one [24], [25]  for instance, have had some success in this respect, but it is probably fair to say that
a deeper  understanding is still needed for phase transitions. \\ 

The regularity properties of the canonical partition function, in the thermodynamic limit, have proved to be an effective method in
answering many of the questions in this field [53]. More recently [55], the 
convexity properties of the thermodynamic potentials have been increasing in prominence toward reaching these goals. Not too surprisingly,
most of the analysis uses the entropic form \ $S_{BGS}$ \ [56], [57] even in cases, such as  systems with long-range interactions, 
when it is far from obvious that \ $S_{BGS}$ \ is, or should be, applicable in describing the collective behaviour of such systems [6]. \\  

Phase transitions involve a substantial re-organisation, and possible re-definition, of the effective degrees of freedom of a system. 
Until now, the tried and tested approach assumes that all phases are described by the same entropy, namely \ $S_{BGS}$. \ 
However, in view of the re-organisation of the system,  it may be conceivable for different entropic functionals to describe different phases. 
To be more concrete, we consider the Tsallis entropy as an example: it depends 
on the non-extensive parameter \ $q$. \ In the above proposal, different values of \ $q$ \ could describe different phases of the same 
system. This should be a sufficient, but not necessary condition. Clearly, using different entropic functionals severely limits the predictable 
power of the proposal, unless one can find a way to somehow connect the microscopic dynamics with the choice of such an entropic functional.
This is essentially ``Boltzmann's program" [54]. Although not proven rigorously, it has been  working successfully for ergodic systems, 
systems with exponential increase in their phase space volume, systems with weak correlations, short-range interactions  etc [54].
As one enhances the classes of  systems under study though, to include systems with strong correlations, long-range interactions, power-law 
increase of the phase space volume, systems out of equilibrium etc,  it is not obvious a  priori that an approach  utilising one functional to 
describe the different phases maybe as successful [6].\\
   
To stay within the group-theoretical framework of the present work, consider for instance a system whose underlying dynamics is given by a 
free group on \ $n$ \  generators \ $\mathcal{F}_n$. \  Such a group has exponential volume growth (18).
Then \ $S_{BGS}$  \ is an effective entropic functional for the description of such a system [12], in analogy with binary systems [7]-[11].
Consider, however a particular phase which is characterised by the addition of more constraints which in other phases would be trivial. 
What we have in mind is the imposition of a non-trivial expectation value of a (space-time) scalar order parameter which results in 
spontaneous symmetry breaking in the Landau-Ginzburg approach to phase transitions [25]. Such an order parameter is usually assumed to 
have modulus zero in the symmetric phase, but its modulus acquires a non-zero vacuum expectation value in the broken symmetry 
phase [24], [25]. As a result, the symmetry group of the broken symmetry phase becomes a subgroup of the set of symmetries of the 
symmetric phase.  Suppose that one imposes such sets of conditions of \ $\mathcal{F}_n$, \ expressed via the relations \ $R$ \
in the group presentation, so that the configuration/phase space of the system is now reduced to a virtually nilpotent group of degree \ $d$. \  
Then the Tsallis entropy with a non-extensive parameter \ $q(d)$ \ may be appropriate for 
describing the thermodynamic behaviour of such a system. At the heart of the matter is extensivity: 
to able to use conventional thermodynamics, the entropy \ $S$ \ of the system of \ $N$ \ degrees of freedom should be extensive, 
namely it should satisfy
  \begin{equation}
      0 \ \neq \ \lim_{N\rightarrow\infty} \frac{S(N)}{N} \ < \infty       
 \end{equation}
 To describe the thermodynamic behaviour of a system then, an entropy functional \ $S$ \ should be used for which (30) is valid. 
 For the case of a nilpotent group, choosing as \ $S$ \ the BGS entropy gives a somewhat trivial result which violates (30), since 
 \begin{equation}
     \lim_{N\rightarrow\infty} \  \frac{\log N}{N} \ = \ 0   
 \end{equation}
   so such a choice could make the thermodynamic analysis of the system problematic. By using the Tsallis entropy
 \ $S_q$ \ for \ $q(d)$, \ for instance, then (30) would be satisfied for a non-zero value of the limit. 
 So two phases of that system could be: one  which is described by \ $\mathcal{F}_n$ \ therefore by the BGS entropy, which 
 is the Tsallis entropy \ $S_q$ \ for \ $q=1$, \ and one which is described by the nilpotent group which is controlled by the Tsallis entropy for 
\  $q(d) \neq 1$. \ Naturally these arguments can also be applied to other entropic functionals 
 having one or more parameters that  may be of interest for Statistical Mechanics [6].\\
 
 It may be worth noticing that we would not expect the converse to be true: it may be possible to describe different phases with the same 
 value of \ $q$ \ in Tsallis entropy. The conventional approach to phase transitions using \ $S_{BGS}$ \  which is \ $S_q$ \ for \ $q=1$ \ is an 
 example. An important question is to address why \ $S_{BGS}$ \ has been so successful in describing phase transitions even though 
 \ $q=1$ \ remains the same for different phases. Within the discrete group-theoretical framework that we consider, the answer appears to be  
 straightforward: \ $S_{BGS}$ \ is associated with the exponential growth rate of (15) as given by (17). This behaviour is typical of free or 
 of  hyperbolic groups which are the majority of groups, as will  be explained in greater detail in the next section. From the viewpoint of 
 thermodynamics, only some aspects of these groups that are non-trivial in the thermodynamic limit, are important. From a geometric 
 viewpoint such features are described as being quasi-isometry invariants of the underlying groups. As mentioned earlier however, 
 any two free groups on \ $m$ \ and \ $n$ \ generators are quasi-isometric to each other even if \ $m\neq n$. \  Hence, even though the systems 
 themselves may be modelled on different groups, their large-scale features are common and both of them can be described by
 \ $S_{BGS}$. \ This idea is similar to the scaling behavior of systems at their critical point: 
 different microscopic systems, having different Hamiltonians can belong to the same universality class as long as they have the same critical 
 exponents and, in the renormalization group language,  can be seen as perturbations of each other by irrelevant operators [53], [54].\\ 
 

\noindent{\large\bf 4.2} \ \ In this discrete group-theoretical context,  it is even possible to explain the ubiquity and success of \ $S_{BGS}$. \  
For this we need the concept of random groups [17], [18] . What the construction of  random groups does, is to assume a presentation 
and try to determine  ``typical" features of the resulting group. To make this precise, one has to specify a way to pick such a group presentation  
and the resulting questions have to be accurately asked within this framework. There are several models of such random groups [17], [18]; 
we will follow the ``density model" [18] since it is sufficiently flexible and still can showcase important pertinent points to the discussion. 
For this, consider the free group on \ $m$ \ generators \ $g_1, \ldots, g_m$, \ indicated by \ $\mathcal{F}_m$ \ as above.  
For any length \ $l\in\mathbb{N}$ \ let \  \ $\partial B (l)$ \ indicate the number of reduced words of length \ $l$ \ in these generators.  
Consider the ``density" parameter \  $0\leq d \leq 1$. \ Pick, uniformly and independently, among the elements of \ $\partial B(l)$, \ a random 
number of \ $(2m-1)^{dl}$ \  relations, collectively indicated by \ $R$. \ A random group at density \ $d$ \ and length \ $l$ \ is the group
 \ $\mathcal{G}$ \ having presentation  \ $\langle g_1, \ldots, g_m \ | \ R \rangle$. \  
A property of \ $\mathcal{G}$ \ occurs almost certainly at density \ $d$, \ if its probability of occurrence
approaches \ 1 \ as \ $l\rightarrow\infty$. \ It may be worth mentioning that in this model of random groups \ $d=0$ \ refers also to the case of 
sub-exponential growth of the number of relations as functions of their length. A fundamental result [18] states that 
if \ $d < 0.5$ \ then the resulting group is almost certainly infinite, hyperbolic, torsion-free (and of cohomological dimension 2). \ If \ $d>0.5$ \
then the resulting group is trivial. Not much is known about the threshold case \ $d=0.5$. \ This result was predated by the proof of [58]
for what turned out to be a special case of the above construction.\\ 

A way to interpret this result from our viewpoint is as follows: Assume the underlying group structure. The kinematics of interest is 
relies on the set of $m$-generators. Consider this to be something like a basis of a Hilbert space, in the case of Quantum Physics.
The allowed set of possibilities, akin to the kinematic Hilbert space, is expressed by the set of words \  $\mathcal{F}_m$ \ in these generators. 
The underlying dynamics is expressed by imposing some constraints and dynamical equations  on the kinematic Hilbert space. 
In the group case, these constraints are provided by the set of relations \ $R$. \ The set of relations has to be sufficiently large so as to
be able to provide sufficient physical solutions of interest. In the above model, the number of relations are substantially smaller than the  
number of possible words of length \ $l$, \ the former being about \ $(2m-1)^{dl}$ \ whereas the latter is \ $(2m_1)^l$. \ However, as noticed
in [18], these numbers are expressed in logarithmic scale, hence the discrepancy between the order of magnitude of words and relations 
encountered in the random group modelled above. Ultimately, someone wishes to take the thermodynamic limit of the underlying system, 
which amounts to taking the word length \ $l\rightarrow\infty$ \ and see what happens in this case. The fact that for \ $d<0.5$ \ 
the underlying group is hyperbolic can be interpreted to mean that in the presence of not too many relations (``under-constrained" system), the 
underlying space is hyperbolic. As such, its volume growth is exponential (17). For the underlying entropy to be extensive (30), it must therefore 
involve a logarithm, namely it should be \ $S_{BGS}$. \ If, on the other hand, the system is ``over-constrained" as for \ $d>0.5$, \ it will have 
too few solutions of  interest, so its thermodynamic behaviour will be trivial. The place where \ $S_q$ \ and other non-extensive entropic forms 
may become of  interest, is in encoding the possible thermodynamic behaviour of the system at the threshold value \ $d=0.5$. \ It is not entirely 
far-fetched to conjecture that  at \ $d=0.5$ \ the volume growth rate increases sub-exponentially or even polynomially. 
In the latter case, as the case of nilpotent groups  indicates, extensivity of the entropy, and its subsequent identification with Clausius' entropy, 
is better  captured by using \ $S_q, \ q \neq 1$ \ instead of the \ $S_{BGS}$ \ functional. 
The thermodynamic behaviour of a system is also insensitive to any changes in a  small number of degrees of 
freedom, especially when the system is far from a phase transition. This is expressed by the quasi-isometry invariance 
of the volume growth function, which in the metric context, guarantees the robustness of the results mentioned in this work, as was also noticed 
in the previous section.\\  
  
Naturally, it is entirely possible that the density model is still too simplistic to describe the thermodynamic behaviour of systems 
described by group-theoretical models, in which case the above arguments either need to be extended, modified  or even abandoned 
altogether.  \\

                                                                       \vspace{5mm}


\centerline{\large\sc 5. \ Conclusions and discussion}

                                                                         \vspace{5mm}

We proposed using discrete groups, which are  finitely generated or even finitely presented, as the kinematical framework for 
building models of systems whose behaviour may be described by a non-additive entropy. We used the Tsallis entropy as a 
paradigmatic case in this work, but most arguments can be extended to more general entropic functionals, obeying similar 
algebraic properties. We stressed, in particular, the significance of nilpotent as well as of hyperbolic groups in this context as well as the 
central role of quasi-isometries in capturing the large-scale properties of the group.    
We also used a specific model of random groups, in an attempt to understand why \ $S_{BGS}$ \ appears to be so widespread 
in modelling of natural systems.\\ 

Motivated by the fundamental result of [42] on the volume growth rate of groups, and combining it with the results of [12], 
we conjectured that a sufficient  condition for phase transitions its to have the different phases described by different values of the 
parameters in the multi-parameter family of functionals that may describe the underlying system.  
We also pointed out why this condition is not necessary. A desirable next step to this present work would be to build concrete 
models of physical interest where the general ideas presented here can be effectively tested in their consistency and physical validity 
and relevance.  \\

                                                                       \vspace{0mm}


                                                        \centerline{\large\sc References}
 
                                                                          \vspace{5mm}
                                                                          
\noindent [1] \ J. Havrda, F. Charv\'{a}t, \ \emph{Kybernetica} {\bf 3}, \ 30 \ (1967).\\
\noindent [2] \ Z. Dar\'{o}czy, \ \emph{Inf. Comp. / Inf. Contr.} {\bf 16}, \ 36 \ (1970).\\
\noindent [3] \ N.A. Cressie, T.R. Read, \ \emph{J. Roy. Stat. Soc. B} {\bf 46}, \ 440 \ (1984).\\
\noindent [4] \ T.R. Read, N.A. Cressie, \ \emph{Goodness of Fit Statistics for Discrete Multivariate Data}, \ Springer,\\ 
                                  \hspace*{6mm} New York \ (1988).\\  
\noindent [5] \ C. Tsallis, \ \emph{J. Stat. Phys.} {\bf 52}, \ 479 \ (1988).\\
\noindent [6] \ C. Tsallis, \ \emph{Introduction to Nonextensive Statistical Mechanics: Approaching a Complex World}\\
                                  \hspace*{6mm}  Springer, \  New York \ (2009).\\ 
 \noindent [7] \ C. Tsallis, M. Gell-Mann, Y. Sato, \ \emph{Proc. Nat. Acad. Sci.} {\bf 102}, \ 15377 \ (2005).\\ 
 \noindent [8] \ A. Rodr\'{i}guez, V. Schw\"{a}mmle, C. Tsallis, \ \emph{J. Stat. Mech.: Theor. Exp.} \ P09006 \ (2008).\\
 \noindent [9] \ R. Hanel, S. Thurner, C. Tsallis, \ \emph{Eur. Phys. J. B} {\bf 72}, \ 263 \ (2009).\\ 
 \noindent [10] \ A. Rodr\'{i}guez, C. Tsallis, \ \emph{J. Math. Phys.} {\bf 51}, \ 073301 \ (2010).\\  
 \noindent [11] \ A. Rodr\'{i}guez, C. Tsallis, \ \emph{J. Math. Phys.} {\bf 53}, \ 023302 \ (2012).\\
 \noindent [12] \ R. Hanel, S. Thurner, \ \emph{Europhys. Lett.} {\bf 96}, \ 50003 \ (2011).\\  
 \noindent [13] \ J.H. Conway, O.D. Friedrichs, D.H. Huson, W.P. Thurston, \ \emph{Beitr\"{a}ge Alg. Geom.} {\bf 42}, \ 475 \\
                                  \hspace*{8mm}  (2001).\\
 \noindent [14] \ R.F. Streater, A.S. Wightman, \ \emph{PCT, Spin \& Statistics and All That}, \ W.A. Benjamin, New \\
                                 \hspace*{8mm} York (1964).\\
 \noindent [15] \ W. Magnus, A. Karass, D. Solitar, \ \emph{Combinatorial Group Theory: Presentations of Groups\\
                                \hspace*{8mm} in Terms of  Generators and Relations}, \ 2nd Ed., \ Dover Publications, \ Mineola NY \ (1976).\\                                  
 \noindent [16] \ R.C. Lyndon, P.E. Schupp, \  \emph{Combinatorial Group Theory}, \ Springer-Verlag, \  Berlin \ (1977).\\ 
 \noindent [17] \ M. Gromov, \ \emph{Hyperbolic groups}, \  \ in \ \ \emph{Essays in group theory}, \ S. Gersten (Ed.), \ MSRI\\
                                \hspace*{8mm}   Publications {\bf 8}, \ Springer, \  New York \  (1987).\\
 \noindent [18] \ M. Gromov, \ \emph{Asymptotic Invariants of Infinite Groups} \  in \ \emph{Geometric Group Theory, Vol.2}, \\
                                \hspace*{8mm}  G.A. Niblo, M.A. Roller (Eds.), \  Cambridge University Press,  \ Cambridge, \ UK \ (1993). \\
 \noindent [19] \ P. de la Harpe, \ \  \emph{Topics in Geometric Group Theory}, \  \ The University of Chicago Press, \\ 
                                 \hspace*{8mm} Chicago \  (2000).\\ 
 \noindent [20] \ J. J. Rotman, \ \ \emph{An Introduction to the Theory of Groups}, \ 4th Edition, \ \ Springer-Verlag, \\
                                \hspace*{8mm}  New York  \ (1995).\\ 
 \noindent [21] \ M. Henneaux, C. Teitelboim, \ \emph{Quantization of Gauge Systems}, \ Princeton University Press, \\
                                \hspace*{8mm} Princeton, NJ \ (1992).\\
 \noindent [22] \ W.P. Thurston, \ \emph{Bull. Amer. Math. Soc.} {\bf 6}, \ 357 \ (1982).\\
 \noindent [23] \ R. M. Wald, \ \emph{General Relativity}, \  University of Chicago Press, \ Chicago \ (1984).\\ 
 \noindent [24] \ C. Itzykson, J.-B. Zuber, \ \emph{Quantum Field Theory}, \ McGraw Hill, \  New York \ (1980).\\
 \noindent [25] \ S. Coleman, \ \emph{Aspects of Symmetry: Selected Erice Lectures}, \ Cambridge University Press, \\
                                \hspace*{8mm}  Cambridge, UK \ (1985).\\
 \noindent [26] \ G. Gallavotti, E.G.D. Cohen, \ \emph{J. Stat. Phys.} {\bf 80}, \ 931 \ (1995).\\ 
 \noindent [27] \ G. Gallavotti, E.G.D. Cohen, \  \emph{Phys. Rev. Lett.} {\bf 74}, \ 2694  \ (1995).\\
 \noindent [28] \ A. Katok, B. Hasselblatt, \ \emph{Introduction to the Modern Theory of Dynamical Systems}, \\ 
                                    \hspace*{8mm} Cambridge University Press, \ Cambridge, UK \ (1995).\\
 \noindent [29] \ V. Efremovich, \ \emph{Usp. Mat. Nauk.} {\bf 8}, \ 189 \ (1953).\\
 \noindent [30] \ A.S. Svar\v{c}, \ \emph{Dokl. Acad. Nauk. SSR} {\bf 105}, \ 32 \ (1955).\\ 
 \noindent [31] \ S. Helgason, \emph{Differential Geometry, Lie Groups and Symmetric Spaces}, \ Academic Press, \\
                                   \hspace*{8mm}  New York \ (1978).\\
 \noindent [32] \ M. Gromov, \ \emph{J. Diff. Geom.} {\bf 13}, \ 231 \ (1978).\\
 \noindent [33] \ E. Ruh, \  \emph{J. Diff. Geom.} {\bf 17}, \ 1 \ (1982).\\
 \noindent [34] \ C. Tsallis, \ \emph{An introduction to nonadditive entropies and a thermostatistical approach of \\
                                    \hspace*{8mm}  inanimate and living matter},  \  \  {\sf arXiv:1403.5425}\\
 \noindent [35] \  J. Dixmier, \ \emph{Publ. Math. I.H.E.S.} {\bf 6}, 13 \ (1960).\\ 
 \noindent [36] \ J.A. Wolf, \ \emph{J. Diff. Geom.} {\bf 2}, \ 421 \ (1968).\\
 \noindent [37] \ J. Milnor, \ \emph{J. Diff. Geom.} {\bf 2}, \ 447 \ (1968).\\
 \noindent [38] \ J. Tits, \ \emph{Jour. Alg.} {\bf 20}, \ 250 \ (1972).\\
 \noindent [39] \ Y. Guivarc'h, \ \emph{C. R. Acad. Sci. Paris, S\'{e}r. A} {\bf 271}, \ 237 \ (1970).\\ 
 \noindent [40] \ Y. Guivarc'h, \ \emph{C. R. Acad. Sci. Paris, S\'{e}r. A} {\bf 272}, \ 1695 \ (1971).\\ 
 \noindent [41] \ H. Bass, \ \emph{Proc. Lond. Math. Soc.} {\bf 25}, \ 603 \ (1972).\\
 \noindent [42] \ M. Gromov, \ \emph{Publ. Math. I.H.E.S.} {\bf 53}, \ 53 \ (1981).\\
 \noindent [43] \ B. Kleiner, \ \emph{J. Amer. Math. Soc.} {\bf 23}, \ 815 \ (2010).\\
 \noindent [44] \ Y. Shalom, T. Tao, \ \emph{Geom. Funct. Anal.} {\bf 20}, \ 1502 \ (2010).\\ 
 \noindent [45] \ P. Pansu, \ \emph{Ergod. Th. \& Dynam. Syst.} {\bf 3}, \ 415, \ (1983).\\
 \noindent [46] \ M. Stoll, \ \emph{J. Lond. Math. Soc.} {\bf 58}, \ 38 \ (1998).\\
 \noindent [47] \ E. Breuillard, E. Le Donne, \ \emph{Proc. Nat. Acad. Sci.} {\bf 110}, \ 19220 \ (2013).\\ 
 \noindent [48] \ B. Farb, \ \emph{Geom. Funct. Anal.} {\bf 8}, \ 810 \ (1998).\\
 \noindent [49] \ D. Osin, \ \emph{Amer. Math. Soc. Mem.}  {\bf 179}, \ 843 \ (2006). \\
 \noindent [50] \ B. Bowditch, \ \emph{Int. J. Alg. Comp.} {\bf 22}, \ 1250016 \ (2012).\\
 \noindent [51] \ A. Manning, \ \emph{Ann. Math.} {\bf 110}, \ 567 \ (1979).\\                      
 \noindent [52] \ G. Besson, G. Courtois, S. Gallot,  \emph{Geom. Funct. Anal.} {\bf 5}, \ 731 \ (1995).\\
 \noindent [53] \ L. Kadanoff, \ \emph{Statistical Physics: Statics, Dynamics and Renormalization}, \ World Scientific, \\ 
                                     \hspace*{8mm} Singapore \ (2000).\\
 \noindent [54] \ G. Gallavotti, \ \  \emph{Statistical Mechanics: A Short Treatise}, \ \ Springer-Verlag, \ Berlin \  (1999).\\
 \noindent [55] \ M. Kastner, \ \emph{Rev. Mod. Phys.} {\bf 80}, 167 \ (2008).\\
 \noindent [56] \ F. Bouchet, S. Gupta, D. Mukamel, \ \emph{Physica A} {\bf 389}, \ 4389 \ (2010).\\
 \noindent [57] \ A. Campa, T. Dauxois, S. Ruffo, \ \emph{Phys. Rep.} {\bf 480}, \ 57 \ (2009).\\ 
 \noindent [58] \ A. Yu. Ol'shanskii, \ \emph{Int. J. Alg. Comp.} {\bf 2}, \ 1 \ (1992).\\

 \end{document}